\documentclass[onecolumn,english,aps,prl,nosuperscriptaddress,notitlepage]{revtex4-1}
\usepackage{mathpazo}
\usepackage[T1]{fontenc}
\usepackage[latin9]{inputenc}
\usepackage{xcolor}
\usepackage{pdfcolmk}
\usepackage{babel}
\usepackage{amsmath}
\usepackage{amssymb}
\usepackage{graphicx}
\PassOptionsToPackage{normalem}{ulem}
\usepackage{ulem}
\usepackage[unicode=true,pdfusetitle,
 bookmarks=false,
 breaklinks=true,pdfborder={0 0 0},pdfborderstyle={},backref=false,colorlinks=true,pdfpagemode=FullScreen]
 {hyperref}
\hypersetup{
 citecolor=blue,filecolor=black,linkcolor=red,urlcolor=blue}

\makeatletter

\providecolor{lyxadded}{rgb}{0,0,1}
\providecolor{lyxdeleted}{rgb}{1,0,0}
\DeclareRobustCommand{\lyxadded}[3]{{\texorpdfstring{\color{lyxadded}{}}{}#3}}

\DeclareRobustCommand{\lyxsout}[1]{\ifx\\#1\else\sout{#1}\fi}

\makeatother

\begin{document}
\title{Spontaneous symmetry breaking induced by interaction in linearly coupled
binary Bose Einstein condensates}
\author{Mateus C. P. dos Santos}
\affiliation{Instituto de Física, Universidade Federal de Goiás, 74.690-970, Goiânia,
Goiás, Brazil}
\author{Wesley B. Cardoso}
\affiliation{Instituto de Física, Universidade Federal de Goiás, 74.690-970, Goiânia,
Goiás, Brazil}
\begin{abstract}
\qquad{}We analyze the spontaneous symmetry breaking (SSB) induced
by one specific component of a linearly coupled binary Bose-Einstein
condensate (BEC). The model is based on linearly coupled Schrödinger
equations with cubic nonlinearity and with a double-well (DW) potential
acting on only one of the atomic components. By numerical simulations,
symmetric and asymmetric ground-states were obtained, and an induced
asymmetry in the partner field was observed. In this sense, we properly
demonstrated that the linear coupling mixing the two-field component
(Rabi coupling) promotes the (in)balance between atomic species, as
well as the appearance of the Josephson and SSB phases.
\end{abstract}
\maketitle

\section{Introduction}

Bose-Einstein condensates (BECs) \citep{ANDERSON_SCIENCE95,DAVIS_PRL95,BRADLEY_PRL95}
offer the opportunity to observe macroscopic quantum effects, providing
a way to study various types of phenomena such as bright \citep{KHAYKOVICH_SCIENCE02,CORNISH_PRL06,MARCHANT_NC13,ABDULLAEV_IJMPB05,SALASNICH_OQL17,STRECKER_NATURE02}
and dark \citep{BURGER_PRL99,DENSCHLAG_SCIENCE00,ANDERSON_PRL01}
solitons, vortices \citep{MATTHEWS_PRL99,MADISON_PRL00}, Anderson
localization of matter waves \citep{BILLY_NATURE08,ROATI_NATURE08},
breathers \citep{DICARLI_PRL19,LUO_PRL20}, self-trapped states supported
by beyond-mean-field interactions (``quantum droplets'') \citep{CABRERA_SCIENCE18,CHEINEY_PRL18,SEMEGHINI_PRL18,DERRICO_PRR19},
etc. In this sense, models with binary bosonic mixtures have been
extensively studied and show a wide application potential. For example,
in Ref. \citep{JAIN_PRA11} by using numerical Quantum Monte Carlo
simulations was demonstrated that a Bose mixture of trapped dipolar
atoms of identical masses and dipole moments presents demixing for
low finite temperatures. Recently, Anderson localization induced by
a linear interaction between the atomic components of binary BECs
(Rabi coupling) was reported in Ref. \citep{dosSANTOS_PRE21}, even
with one of the components not being directly influenced by the disordered
potential.

Another important phenomenon observed in binary bosonic mixtures is
the symmetry breaking. Indeed, the stability \citep{XU_PRA08} and
dynamic tunneling properties of binary BECs trapped by double-well
potential were investigated in Ref. \citep{SATIJA_PRA09}. Specifically,
in \citep{MAZZARELLA_JPB09}, the authors showed the possibility of
describing the imbalance between atomic populations through effective
equations, analogous to the dynamic equations of the coupled pendulum.
Spontaneous symmetry breaking (SSB) was also studied in a two-component
model in double-well nonlinear (pseudo)potentials \citep{ACUS_PD12}.
In this case, cubic nonlinear Schrödinger equations with nonlinear
coupling were considered, so that the nonlinear modulation function
acts in the system as a double-well potential, producing asymmetric
profiles under certain conditions. Moreover, SSB was found in some
other models such as Bose-Fermi mixtures trapped by double-well potential
\citep{ADHIKARI_PRA10}, two-component linearly coupled system with
the intrinsic cubic nonlinearity and the harmonic-oscillator confining
potential (valid for BECs and optical systems) \citep{HACKER_SYMMETRY21},
linearly coupled Korteweg-de Vries systems \citep{ESPINOSA-CERON_CHAOS12},
etc.

The description of ultra-cold interacting bosonic gases based on mean-field
approximations is dictated by a well-known time-dependent three-dimensional
Gross-Pitaevskii equations (GPEs) \citep{Pitaevskii_03,PETHICK_08}.
Therefore, for BECs strongly trapped in the transverse direction,
it is possible to apply a dimensional reduction method capable of
producing an effective equation that describes the dynamics of the
system through a quasi-one-dimensional equation for the longitudinal
component of the wave function, while keeping the transversal one
``frozen'' \citep{SALASNICH_PRA02,MASSIGNAN_PRA03,BUITRAGO_JPB09,Mateo_PRA08,Couto_AP18}.
In particular, Salasnich \emph{et. al.} \citep{SALASNICH_PRA02} presented
a time-dependent nonpolynomial Schrödinger equation that accurately
described an anisotropic BEC confined strongly in the transverse direction
by a harmonic trap (cigar shaped), obtained through a variational
approach. This technique has been widely used for several other models
\citep{SALASNICH_PRA02_2,SALASNICH_PRA06,SALASNICH_PRA07,ADHIKARI_NJP09,Cardoso_PRE11,dosSantos_PLA19,Santos_JPB19,dosSANTOS_EPJST21}.

The goal of the present work is to carry out a systematic study about
the symmetry breaking induced by a specific atomic component of a
linearly coupled binary BEC. To this end, differently from the previous
studies, we will analyze the behavior of a specific atomic component
that interacts with a second component (partner) via Rabi coupling,
but it is not directly in contact with the double-well potential.
The presence of a double-well external potential confining a single
component induces an asymmetry in the balance between atomic populations
as well as in the profile of each component. The rest of the paper
is organized as follows. In the Sec. \ref{Sec2}, we present the theoretical
model. The numerical simulations and our analysis are shown in Sec.
\ref{Sec3}. We summarize the paper in Sec. \ref{Sec4}.

\section{Theoretical model \label{Sec2}}

We consider a diluted BEC of two components, close to absolute zero
temperature, whose atomic species interact linearly. This type of
interaction is obtained when the interspecies interaction of the binary
BEC with Rabi-coupling is turned off, remaining only the linear component.
Here, we are considering that with the auxiliary laser field, only
one of the atomic species (component 1) is in ``direct contact'' with
a double-well type potential.

In systems where the three-dimensional external potential is strongly
confining in the transverse direction ($y,z$), it is convenient to
use a dimensional reduction procedure, making it possible to use one-dimensional
dynamic equations to describe the behavior of the binary BEC \citep{Young_PRA10,Adhikari_JPB11}.
Therefore, the effective coupled equations that precisely describes
this model can be written as

\begin{subequations}
\begin{equation}
i\frac{\partial f_{1}}{\partial t}=-\frac{1}{2}\frac{\partial^{2}f_{1}}{\partial x^{2}}+V_{DW}(x)f_{1}+\alpha|f_{1}|^{2}f_{1}+\kappa f_{2},\label{EQ0}
\end{equation}
\begin{equation}
i\frac{\partial f_{2}}{\partial t}=-\frac{1}{2}\frac{\partial^{2}f_{2}}{\partial x^{2}}+\gamma|f_{2}|^{2}f_{2}+\kappa f_{1},\label{EQ1}
\end{equation}
\end{subequations} where $f_{j}(x,t)$, ($j=1$,$2$) is the macroscopic
wave function of the binary BEC, $\alpha$ and $\gamma$ are the strengths
of intraspecies interactions relative to atomic species $1$ and $2$,
respectively, and $\kappa$ is the linear coupling constant that effectively
mediates the interaction between the two components of bosonic gas.
The system is subject to joint normalization condition 
\begin{equation}
\int_{-\infty}^{+\infty}(|f_{1}|^{2}+|f_{2}|^{2})dx=N_{1}+N_{2},\label{NORM}
\end{equation}
where $N_{j}(j=1,2)$ is the individual norm of each atomic component.
In the results below the fields $f_{1}$ and $f_{2}$ are initially
normalized to unity, resulting in the joint normalization $N_{1}+N_{2}=2$.

In Ref. \citep{HACKER_SYMMETRY21}, the dynamics of linearly coupled
two-component systems with intrinsic cubic nonlinearity and the confinement
potential of the harmonic oscillator (HO) acting on only one of the
components was studied. It was shown through approximate analytics
solutions of ground state, first exited state (dipole mode) and numerical
solutions that this asymmetric model (half-trapped system) has SSB,
presenting imbalance between the norms of the components. Differently
from the Ref. \citep{HACKER_SYMMETRY21}, in the present study we
consider that the component 1 is subject to the action of a double-well
(DW) potential $V_{DW}(x)$, given by

\begin{equation}
V_{DW}=-V_{0}\left(\frac{1}{\cosh(x-x_{0})^{2}}+\frac{1}{\cosh(x+x_{0})^{2}}\right),\label{DW}
\end{equation}
obtained from the combination of two Pöschl--Teller (PT) potentials
centered at $\pm x_{0}$, and separated by a barrier of height $V_{0}$.
The single PT potential have been studied in systems described by
the Schrödinger equation and their eigenvalues and eigenfunctions
are known \citep{landau2013quantum,Posh_ZP33}. Effects such as the
absence of reflection of the Gaussian wave packet scattered by a PT
well \citep{Kiriushcheva_AJP98,Lekner_AJP07}, second and third harmonic
generation \citep{Sakiroglu_PLA12} and intersubband absorption in
optical systems \citep{Radovanovic_PLA00,Yildirim_PRB05} were reported
too.

In Refs. \citep{MAZZARELLA_JPB09,MAZZARELA_JPB10}, the DW potential
(\ref{DW}) was used to study the atomic Josephson effect in binary
BECs with nonlinear interaction. The system was modeled by two coupled
GPE, confined by a double-well potential along the axial direction
and a strong harmonic confinement in the transverse direction, where
effective equations describe the dynamic behavior of atomic components.
Also, in Ref. \citep{Mazzarella_PRA10} was reported the existence
of SSB in a single BEC component trapped by a double-PT potential
(similar to the one we are using here, i.e., Eq. (\ref{DW})). In
this case, the dimensional reduction process \citep{SALASNICH_PRA02}
was employed to produce a nonpolynomial Schrödinger equation (1D)
capable of accurately describing the longitudinal dynamics of the
three-dimensional model, generating (a)symmetrical and collapsed ground-states.
Numerical results are presented in the next section.

\section{Numerical results \label{Sec3}}

In this section, we present the numerical results obtained from the
integration of Eqs. (\ref{EQ0}) and (\ref{EQ1}), under the joint
normalization condition (\ref{NORM}). With purpose of producing stationary
states, whose configuration has the lowest energy (ground states),
we use an imaginary-time Runge-Kutta algorithm based on the Fourier
spectral method \citep{Yang_10}.

The ground states presented below are obtained from the slightly asymmetrical
initial condition

\begin{equation}
f(x,0)=\delta_{-}\exp\left[\frac{(x-x_{0})^{2}}{2}\right]+\delta_{+}\exp\left[\frac{(x+x_{0})^{2}}{2}\right],\label{IC}
\end{equation}
with $\delta_{+}=1.01$ and $\delta_{-}=1$. It is worth mentioning
that in cases where the asymmetry is placed in the direction of the
right well (i.e., $\delta_{+}=1.0$ and $\delta_{-}=1.01$) the results
are similar, but with $x\rightarrow-x$.

The ground states obtained from (\ref{IC}), by considering the DW
potential (\ref{DW}) and on the joint normalization (\ref{NORM}),
are of two types: the symmetric (Josephson phase) and those with broken
symmetry (SSB phase). Symmetric profiles have equally distributed
energy density between the wells, differently from the asymmetric
ones.

\begin{figure*}[t]
\centering \includegraphics[width=0.8\textwidth]{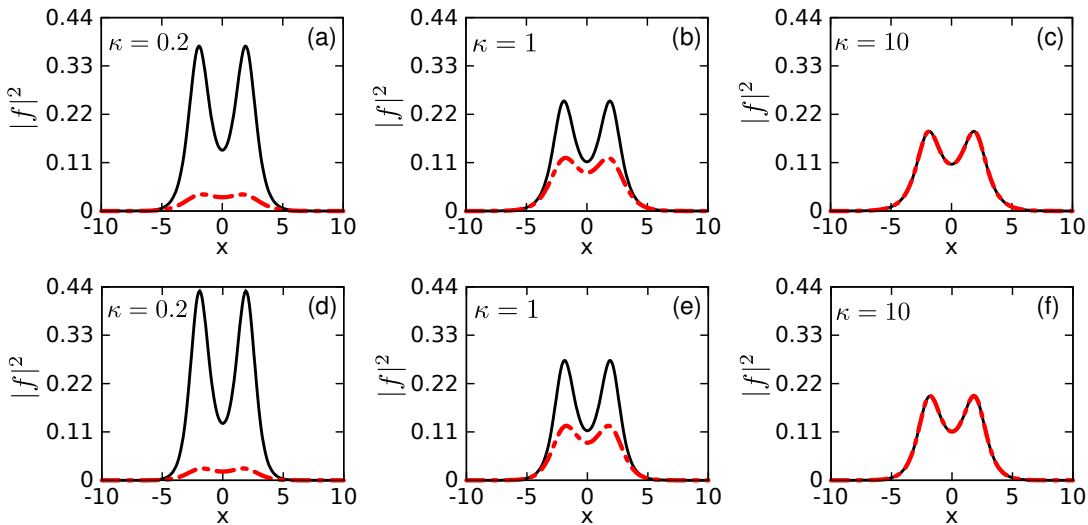} 
\caption{Density profiles $|f_{1}(x)|^{2}$ and $|f_{2}(x)|^{2}$ versus $x$
for a binary BEC subject to the normalization condition (\ref{NORM}).
The values of the nonlinearity parameters used here are (a-c) $\alpha=\gamma=0.5$
and (d-f) $\alpha=\gamma=-0.1$. The profiles $|f_{1}(x)|^{2}$ and
$|f_{2}(x)|^{2}$ are plotted in black solid lines and red dash-dotted
lines, respectively. The other parameters are $V_{0}=1$ and $x_{0}=2$.}
\label{F1}
\end{figure*}
\begin{figure*}[t]
\centering \includegraphics[width=0.8\textwidth]{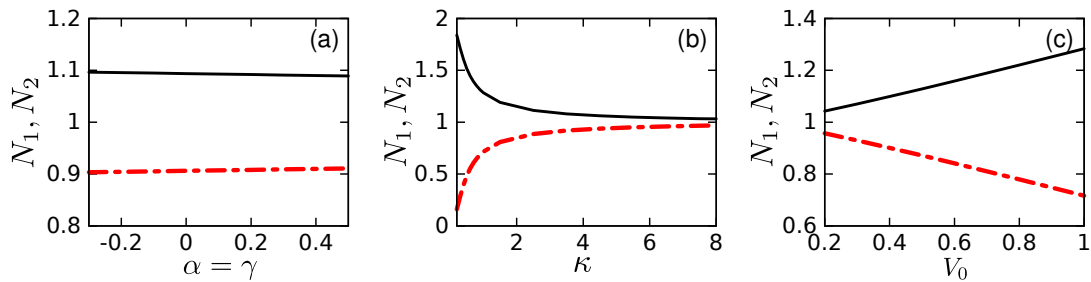} 
\caption{Comparison between the value of the norms $N_{1}$ and $N_{2}$, obtained
via profiles $f_{1}(x)$ and $f_{2}(x)$, versus the parameters (a)
$\alpha=\gamma$, (b) $\kappa$ and (c) $V_{0}$. The norms $N_{1}$
and $N_{2}$ are displayed in solid (black) lines and dash-dotted
(red) lines, respectively. The parameters used here are (a) $\kappa=3$
and $V_{0}=1$; (b) are $\alpha=\gamma=-0.1$ and $V_{0}=1$; (c)
the same as in (b) but now considering $\kappa=1$.}
\label{F1-1}
\end{figure*}
We started our analysis by studying the influence of the coupling
on localization. In Fig. \ref{F1}(a-c) we present symmetric ground
state profiles for the case of self-repulsive binary BEC. Under these
conditions, i.e., $\alpha=\gamma>0$, the field $f_{2}$ is delocalized
in the absence of interaction between the components ($\kappa=0$).
On the other hand, for a weak coupling (see Fig. \ref{F1}(a)) we
verify that there is a localization of component 2, indicating that
field $f_{1}$ acts on field $f_{2}$ simulating an external trap.
It is important to note that the shapes of the densities are quite
different, with $|f_{1}|^{2}$ having a depth much greater than the
density $|f_{2}|^{2}$. This behavior disappears when increasing the
coupling intensity, as can be seen in Fig. \ref{F1}(b-c). Specifically\lyxadded{Wesley B Cardoso}{Fri Feb  4 18:34:59 2022}{,}
in Fig. \ref{F1}(c), with strong coupling $\kappa=10$, the fields
have practically the same configuration. Similar results are found
in the weak self-attractive regime. In this case the self-attractive
interaction ($\alpha,\gamma<0$) produces densities with peaks slightly
higher than the repulsive cases, as observed by Fig. \ref{F1}(d-f).

In general, we observe that a systematic increase in coupling strength
produces a unification of densities, e.g., causing the number of condensed
atoms to behave practically equal with each other, as well as the
shape of fields $f_{1}$ and $f_{2}$. In order to study in detail
this behavior, we analyzed the influence of parameters $\alpha$,
$\gamma$, $\kappa$, and $V_{0}$ on the value of the norms of components
1 and 2. We observe that the interspecies interaction, both attractive
and repulsive, do not significantly change the norm of the components,
as can be seen in Fig. \ref{F1-1}(a). On the other hand, the interspecies
interaction parameter ($\kappa$) has great importance in the individual
norm values. The Fig. \ref{F1-1}(b) shows that the binary BEC is
sensitive to the increment of the linear coupling strength. For example,
with $\kappa=0.2$, the difference between the individual norms $N_{1}-N_{2}=1.68$,
while for $\kappa=7.5$, the difference reduces to $N_{1}-N_{2}=0.07$.
The difference between the individual norms presents a nonlinear pattern,
which slowly goes to zero with increasing of $\kappa$. From a phenomenological
point of view, this behavior makes the condensed cloud composed of
component 2 equal, both in shape and in the number of condensed atoms,
to component 1, in which the latter is under influence of the DW potential
(see Fig. \ref{F1}(c,f)). The height of the barrier also modifies
the relationship between individual norms. However, unlike the previous
case, the increment of $V_{0}$ promotes the linear increment of the
difference $N_{1}-N_{2}$, as shown in Fig. \ref{F1-1}(c).

\begin{figure*}[t]
\centering \includegraphics[width=0.8\textwidth]{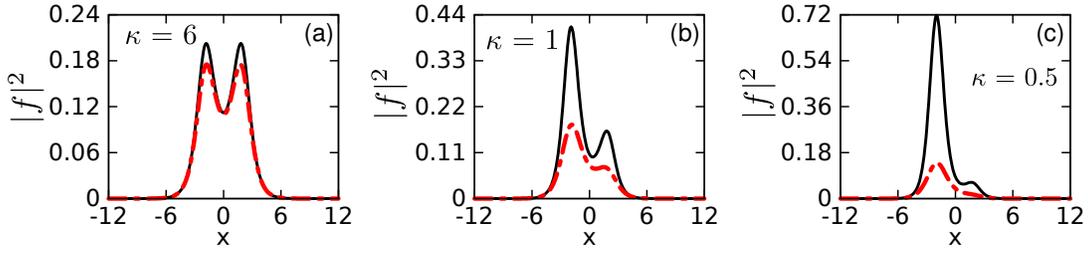} \caption{Density profiles $|f_{1}(x)|^{2}$ and $|f_{2}(x)|^{2}$ versus $x$
for an attractive binary BEC subject to the normalization condition
(\ref{NORM}). The profiles $|f_{1}(x)|^{2}$ and $|f_{2}(x)|^{2}$
are displayed in solid (black) lines and dash-dotted (red) lines,
respectively. The other parameter values used here were $\alpha=\gamma=-0.3$,
$V_{0}=1$ and $x_{0}=2$.}
\label{F3-2}
\end{figure*}

All results presented above were obtained in the self-repulsive or
weak self-attractive regime. In these configurations, the profiles
have a symmetrical shape $|f_{1,2}(x)|^{2}=|f_{1,2}(-x)|^{2}$. However,
for certain parameter values (detailed below) the symmetry of the
profiles is broken (SSB phase) and they start to present a particle
density displaced from the center of the trap. A typical example of
the appearance of asymmetric profiles is shown in Fig. \ref{F3-2}.
In the strong coupling regime (Fig. \ref{F3-2}(a)), for example $\kappa=6$,
the densities $|f_{1}|^{2}$ and $|f_{2}|^{2}$ have a symmetrical
shape. However, by decreasing the coupling strength, the ground states
of both fields start to present broken symmetry, as can be seen in
Fig. \ref{F3-2}(b,c). In these cases, we observe that the left hump
is larger than the right hump for both atomic components. It is noteworthy
that the difference between the amplitude of the humps increases by
decreasing the coupling intensity, resulting in the limit case for
which the profiles have only one hump located on the left with maximum
amplitude in both components.

\begin{figure}[tb]
\centering \includegraphics[width=1\columnwidth]{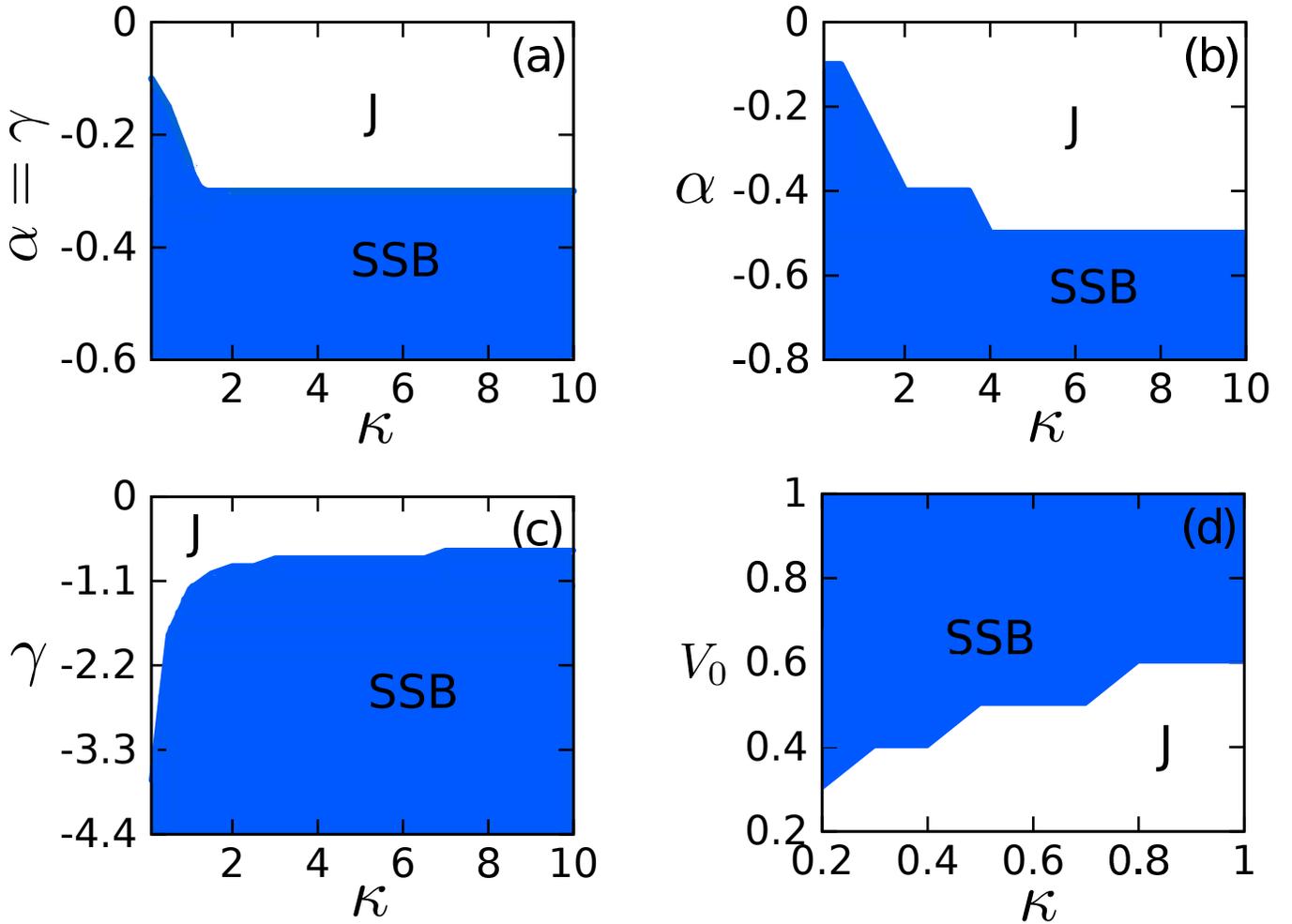}

\caption{Phase diagrams of the attractive BEC in the symmetric double-well
potential $V(x)$ with two minima centered at $\pm2$. The diagrams
represents (a) $\kappa$ versus $\alpha=\gamma$, with $V_{0}=1$;
(b) $\kappa$ versus $\alpha$, with $V_{0}=1$ and $\gamma=0$; (c)
$\kappa$ versus $\gamma$, with $V_{0}=1$ and $\alpha=0$; and (d)
$\kappa$ versus $V_{0}$, with $\alpha=\gamma=-0.4$. The Josephson
phase (J) is represented by the empty area while the spontaneous symmetry
breaking (SSB) is represented by the filled area (blue).}

\label{F3}
\end{figure}

In the previous analyses considering the self-attractive regime, we
observed that binary BECs with $\alpha=\gamma=-0.3$ present symmetrical
profiles for $\kappa\geq1.2$. For the same setting, but with $\kappa<1.2$,
the ground states change abruptly, starting to present an asymmetric
profile. Therefore, we carried out a systematic study to analyze the
regions of existence of symmetric and asymmetric profiles by changing
the values of the parameters $\alpha,\gamma,V_{0}$ and $\kappa$.
The existence diagram of the SSB and Josephson phases, with respect
to the values of the intra- and inter-species interaction parameters
($\alpha=\gamma,\kappa$), shows that by increasing the coupling strength
will generate symmetrical solutions. However, for $\alpha=\gamma<-0.3$
only asymmetric ground states are found, as can be seen in Fig. \ref{F3}(a).
A similar behavior is observed when fixing the auto-interaction parameter
of component 2 ($\gamma=0$) and varying $\alpha$ and $\kappa$ (see
Fig. \ref{F3}(b)). Then, for a strong coupling regime, a self-attracting
interaction $\alpha<-0.5$ is necessary to produce asymmetric profiles.

In Fig. \ref{F3}(c) we verify the influence of the interspecies interaction
of component 2 in the absence of auto-interaction in component 1 (i.e.,
$\alpha=0$), on the shape of the ground states. For example, by considering
a strong coupling regime with $\kappa=10$, asymmetric profiles are
obtained for $\gamma<-1.1$. However, under a weak coupling ($\kappa=0.2$),
the same behavior appears for $\gamma<-3.3$. We highlight that this
behavior is contrary to those observed above. Indeed, in general the
system is less sensitive to $\gamma$ variation, which can be observed
by the area of Josephson phase (J) of the diagram $(\gamma,\kappa)$.
Finally, we analyze the behavior of the ground states in relation
to the coupling parameter and height of the DW barrier potential.
Fig. \ref{F3}(d) shows that increasing the height of the barrier
leads to symmetry breaking. Conversely, by setting $V_{0}$, the coupling
increment tends to generate symmetrical profiles. For example, choosing
$\alpha=\gamma<-0.4$, only asymmetric states are obtained in the
region $0.6<V_{0}<1$ and $0.2<\kappa<1$.

\section{Conclusion \label{Sec4}}

In conclusion, by considering a binary Bose-Einstein condensate with
linear interaction between species, we investigated the existence
of induced symmetry breaking. In this model, only one atomic component
is under the influence of a double well potential, which for certain
parameters induces the appearance of asymmetric ground states both
in the actuated field ($f_{1}$) and in the partner field ($f_{2}$).
Thus, we observe that component 1 acts as a confining potential on
component 2, simulating an asymmetric double-well potential for certain
parameter values. These added effects promote symmetry breaking (SSB
phase) in both atomic components, where the resulting densities have
center of mass displaced in relation to the center of the trap. Symmetrical
profiles (Josephson phase) were also found. They are very sensitive
to variation in intraspecies interaction, as well as the intensity
of the double-well potential when individual norms are considered.
As observed here, the coupling strength drastically influences the
phase of the ground state profiles obtained in the self-attractive
regime ($\alpha,\gamma<0$). Finally, diagrams were used to present
the region of existence of both phases for different parameters. This
work intends to advance the study of binary BECs, but also sheds light
on the investigation of induction effects in partner fields for other
types of coupled systems.

\section*{Acknowledgments}

The author acknowledges the financial support of the Brazilian agencies
CNPq (\#306065/2019-3 \& \#425718/2018-2), CAPES, and FAPEG (PRONEM
\#201710267000540 \& PRONEX \#201710267000503). This work was also
performed as part of the Brazilian National Institute of Science and
Technology (INCT) for Quantum Information (\#465469/2014-0). WBC also
thanks Juracy Leandro dos Santos for his contribution in implementing
part of the infrastructure used in the simulation processes.

\bibliographystyle{apsrev4-2}
\bibliography{REF_SSB_Two_Fields}

\begin{thebibliography}{58}%
\makeatletter
\providecommand \@ifxundefined [1]{%
 \@ifx{#1\undefined}
}%
\providecommand \@ifnum [1]{%
 \ifnum #1\expandafter \@firstoftwo
 \else \expandafter \@secondoftwo
 \fi
}%
\providecommand \@ifx [1]{%
 \ifx #1\expandafter \@firstoftwo
 \else \expandafter \@secondoftwo
 \fi
}%
\providecommand \natexlab [1]{#1}%
\providecommand \enquote  [1]{``#1''}%
\providecommand \bibnamefont  [1]{#1}%
\providecommand \bibfnamefont [1]{#1}%
\providecommand \citenamefont [1]{#1}%
\providecommand \href@noop [0]{\@secondoftwo}%
\providecommand \href [0]{\begingroup \@sanitize@url \@href}%
\providecommand \@href[1]{\@@startlink{#1}\@@href}%
\providecommand \@@href[1]{\endgroup#1\@@endlink}%
\providecommand \@sanitize@url [0]{\catcode `\\12\catcode `\$12\catcode
  `\&12\catcode `\#12\catcode `\^12\catcode `\_12\catcode `\%12\relax}%
\providecommand \@@startlink[1]{}%
\providecommand \@@endlink[0]{}%
\providecommand \url  [0]{\begingroup\@sanitize@url \@url }%
\providecommand \@url [1]{\endgroup\@href {#1}{\urlprefix }}%
\providecommand \urlprefix  [0]{URL }%
\providecommand \Eprint [0]{\href }%
\providecommand \doibase [0]{https://doi.org/}%
\providecommand \selectlanguage [0]{\@gobble}%
\providecommand \bibinfo  [0]{\@secondoftwo}%
\providecommand \bibfield  [0]{\@secondoftwo}%
\providecommand \translation [1]{[#1]}%
\providecommand \BibitemOpen [0]{}%
\providecommand \bibitemStop [0]{}%
\providecommand \bibitemNoStop [0]{.\EOS\space}%
\providecommand \EOS [0]{\spacefactor3000\relax}%
\providecommand \BibitemShut  [1]{\csname bibitem#1\endcsname}%
\let\auto@bib@innerbib\@empty
\bibitem [{\citenamefont {Anderson}\ \emph {et~al.}(1995)\citenamefont
  {Anderson}, \citenamefont {Ensher}, \citenamefont {Matthews}, \citenamefont
  {Wieman},\ and\ \citenamefont {Cornell}}]{ANDERSON_SCIENCE95}%
  \BibitemOpen
  \bibfield  {author} {\bibinfo {author} {\bibfnamefont {M.~H.}\ \bibnamefont
  {Anderson}}, \bibinfo {author} {\bibfnamefont {J.~R.}\ \bibnamefont
  {Ensher}}, \bibinfo {author} {\bibfnamefont {M.~R.}\ \bibnamefont
  {Matthews}}, \bibinfo {author} {\bibfnamefont {C.~E.}\ \bibnamefont
  {Wieman}},\ and\ \bibinfo {author} {\bibfnamefont {E.~A.}\ \bibnamefont
  {Cornell}},\ }\href {https://doi.org/10.1126/science.269.5221.198} {\bibfield
   {journal} {\bibinfo  {journal} {Science}\ }\textbf {\bibinfo {volume}
  {269}},\ \bibinfo {pages} {198} (\bibinfo {year} {1995})}\BibitemShut
  {NoStop}%
\bibitem [{\citenamefont {Davis}\ \emph {et~al.}(1995)\citenamefont {Davis},
  \citenamefont {Mewes}, \citenamefont {Andrews}, \citenamefont {van Druten},
  \citenamefont {Durfee}, \citenamefont {Kurn},\ and\ \citenamefont
  {Ketterle}}]{DAVIS_PRL95}%
  \BibitemOpen
  \bibfield  {author} {\bibinfo {author} {\bibfnamefont {K.~B.}\ \bibnamefont
  {Davis}}, \bibinfo {author} {\bibfnamefont {M.~O.}\ \bibnamefont {Mewes}},
  \bibinfo {author} {\bibfnamefont {M.~R.}\ \bibnamefont {Andrews}}, \bibinfo
  {author} {\bibfnamefont {N.~J.}\ \bibnamefont {van Druten}}, \bibinfo
  {author} {\bibfnamefont {D.~S.}\ \bibnamefont {Durfee}}, \bibinfo {author}
  {\bibfnamefont {D.~M.}\ \bibnamefont {Kurn}},\ and\ \bibinfo {author}
  {\bibfnamefont {W.}~\bibnamefont {Ketterle}},\ }\href
  {https://doi.org/10.1103/PhysRevLett.75.3969} {\bibfield  {journal} {\bibinfo
   {journal} {Phys. Rev. Lett.}\ }\textbf {\bibinfo {volume} {75}},\ \bibinfo
  {pages} {3969} (\bibinfo {year} {1995})}\BibitemShut {NoStop}%
\bibitem [{\citenamefont {Bradley}\ \emph {et~al.}(1995)\citenamefont
  {Bradley}, \citenamefont {Sackett}, \citenamefont {Tollett},\ and\
  \citenamefont {Hulet}}]{BRADLEY_PRL95}%
  \BibitemOpen
  \bibfield  {author} {\bibinfo {author} {\bibfnamefont {C.~C.}\ \bibnamefont
  {Bradley}}, \bibinfo {author} {\bibfnamefont {C.~A.}\ \bibnamefont
  {Sackett}}, \bibinfo {author} {\bibfnamefont {J.~J.}\ \bibnamefont
  {Tollett}},\ and\ \bibinfo {author} {\bibfnamefont {R.~G.}\ \bibnamefont
  {Hulet}},\ }\href {https://doi.org/10.1103/PhysRevLett.75.1687} {\bibfield
  {journal} {\bibinfo  {journal} {Phys. Rev. Lett.}\ }\textbf {\bibinfo
  {volume} {75}},\ \bibinfo {pages} {1687} (\bibinfo {year}
  {1995})}\BibitemShut {NoStop}%
\bibitem [{\citenamefont {Khaykovich}\ \emph {et~al.}(2002)\citenamefont
  {Khaykovich}, \citenamefont {Schreck}, \citenamefont {Ferrari}, \citenamefont
  {Bourdel}, \citenamefont {Cubizolles}, \citenamefont {Carr}, \citenamefont
  {Castin},\ and\ \citenamefont {Salomon}}]{KHAYKOVICH_SCIENCE02}%
  \BibitemOpen
  \bibfield  {author} {\bibinfo {author} {\bibfnamefont {L.}~\bibnamefont
  {Khaykovich}}, \bibinfo {author} {\bibfnamefont {F.}~\bibnamefont {Schreck}},
  \bibinfo {author} {\bibfnamefont {G.}~\bibnamefont {Ferrari}}, \bibinfo
  {author} {\bibfnamefont {T.}~\bibnamefont {Bourdel}}, \bibinfo {author}
  {\bibfnamefont {J.}~\bibnamefont {Cubizolles}}, \bibinfo {author}
  {\bibfnamefont {L.~D.}\ \bibnamefont {Carr}}, \bibinfo {author}
  {\bibfnamefont {Y.}~\bibnamefont {Castin}},\ and\ \bibinfo {author}
  {\bibfnamefont {C.}~\bibnamefont {Salomon}},\ }\href
  {https://doi.org/10.1126/science.1071021} {\bibfield  {journal} {\bibinfo
  {journal} {Science}\ }\textbf {\bibinfo {volume} {296}},\ \bibinfo {pages}
  {1290} (\bibinfo {year} {2002})}\BibitemShut {NoStop}%
\bibitem [{\citenamefont {Cornish}\ \emph {et~al.}(2006)\citenamefont
  {Cornish}, \citenamefont {Thompson},\ and\ \citenamefont
  {Wieman}}]{CORNISH_PRL06}%
  \BibitemOpen
  \bibfield  {author} {\bibinfo {author} {\bibfnamefont {S.~L.}\ \bibnamefont
  {Cornish}}, \bibinfo {author} {\bibfnamefont {S.~T.}\ \bibnamefont
  {Thompson}},\ and\ \bibinfo {author} {\bibfnamefont {C.~E.}\ \bibnamefont
  {Wieman}},\ }\href {https://doi.org/10.1103/PhysRevLett.96.170401} {\bibfield
   {journal} {\bibinfo  {journal} {Phys. Rev. Lett.}\ }\textbf {\bibinfo
  {volume} {96}},\ \bibinfo {pages} {170401} (\bibinfo {year}
  {2006})}\BibitemShut {NoStop}%
\bibitem [{\citenamefont {Marchant}\ \emph {et~al.}(2013)\citenamefont
  {Marchant}, \citenamefont {Billam}, \citenamefont {Wiles}, \citenamefont
  {Yu}, \citenamefont {Gardiner},\ and\ \citenamefont
  {Cornish}}]{MARCHANT_NC13}%
  \BibitemOpen
  \bibfield  {author} {\bibinfo {author} {\bibfnamefont {A.~L.}\ \bibnamefont
  {Marchant}}, \bibinfo {author} {\bibfnamefont {T.~P.}\ \bibnamefont
  {Billam}}, \bibinfo {author} {\bibfnamefont {T.~P.}\ \bibnamefont {Wiles}},
  \bibinfo {author} {\bibfnamefont {M.~M.~H.}\ \bibnamefont {Yu}}, \bibinfo
  {author} {\bibfnamefont {S.~A.}\ \bibnamefont {Gardiner}},\ and\ \bibinfo
  {author} {\bibfnamefont {S.~L.}\ \bibnamefont {Cornish}},\ }\href
  {https://doi.org/10.1038/ncomms2893} {\bibfield  {journal} {\bibinfo
  {journal} {Nat. Commun.}\ }\textbf {\bibinfo {volume} {4}},\ \bibinfo {pages}
  {1865} (\bibinfo {year} {2013})}\BibitemShut {NoStop}%
\bibitem [{\citenamefont {{Kh. Abdullaev}}\ \emph {et~al.}(2005)\citenamefont
  {{Kh. Abdullaev}}, \citenamefont {Gammal}, \citenamefont {Kamchatnov},\ and\
  \citenamefont {Tomio}}]{ABDULLAEV_IJMPB05}%
  \BibitemOpen
  \bibfield  {author} {\bibinfo {author} {\bibfnamefont {F.}~\bibnamefont {{Kh.
  Abdullaev}}}, \bibinfo {author} {\bibfnamefont {A.}~\bibnamefont {Gammal}},
  \bibinfo {author} {\bibfnamefont {A.~M.}\ \bibnamefont {Kamchatnov}},\ and\
  \bibinfo {author} {\bibfnamefont {L.}~\bibnamefont {Tomio}},\ }\href
  {https://doi.org/10.1142/S0217979205032279} {\bibfield  {journal} {\bibinfo
  {journal} {Int. J. Mod. Phys. B}\ }\textbf {\bibinfo {volume} {19}},\
  \bibinfo {pages} {3415} (\bibinfo {year} {2005})}\BibitemShut {NoStop}%
\bibitem [{\citenamefont {Salasnich}(2017)}]{SALASNICH_OQL17}%
  \BibitemOpen
  \bibfield  {author} {\bibinfo {author} {\bibfnamefont {L.}~\bibnamefont
  {Salasnich}},\ }\href {https://doi.org/10.1007/s11082-017-1247-5} {\bibfield
  {journal} {\bibinfo  {journal} {Opt. Quantum Electron.}\ }\textbf {\bibinfo
  {volume} {49}},\ \bibinfo {pages} {409} (\bibinfo {year} {2017})}\BibitemShut
  {NoStop}%
\bibitem [{\citenamefont {Strecker}\ \emph {et~al.}(2002)\citenamefont
  {Strecker}, \citenamefont {Partridge}, \citenamefont {Truscott},\ and\
  \citenamefont {Hulet}}]{STRECKER_NATURE02}%
  \BibitemOpen
  \bibfield  {author} {\bibinfo {author} {\bibfnamefont {K.~E.}\ \bibnamefont
  {Strecker}}, \bibinfo {author} {\bibfnamefont {G.~B.}\ \bibnamefont
  {Partridge}}, \bibinfo {author} {\bibfnamefont {A.~G.}\ \bibnamefont
  {Truscott}},\ and\ \bibinfo {author} {\bibfnamefont {R.~G.}\ \bibnamefont
  {Hulet}},\ }\href {https://doi.org/10.1038/nature747} {\bibfield  {journal}
  {\bibinfo  {journal} {Nature}\ }\textbf {\bibinfo {volume} {417}},\ \bibinfo
  {pages} {150} (\bibinfo {year} {2002})}\BibitemShut {NoStop}%
\bibitem [{\citenamefont {Burger}\ \emph {et~al.}(1999)\citenamefont {Burger},
  \citenamefont {Bongs}, \citenamefont {Dettmer}, \citenamefont {Ertmer},
  \citenamefont {Sengstock}, \citenamefont {Sanpera}, \citenamefont
  {Shlyapnikov},\ and\ \citenamefont {Lewenstein}}]{BURGER_PRL99}%
  \BibitemOpen
  \bibfield  {author} {\bibinfo {author} {\bibfnamefont {S.}~\bibnamefont
  {Burger}}, \bibinfo {author} {\bibfnamefont {K.}~\bibnamefont {Bongs}},
  \bibinfo {author} {\bibfnamefont {S.}~\bibnamefont {Dettmer}}, \bibinfo
  {author} {\bibfnamefont {W.}~\bibnamefont {Ertmer}}, \bibinfo {author}
  {\bibfnamefont {K.}~\bibnamefont {Sengstock}}, \bibinfo {author}
  {\bibfnamefont {A.}~\bibnamefont {Sanpera}}, \bibinfo {author} {\bibfnamefont
  {G.~V.}\ \bibnamefont {Shlyapnikov}},\ and\ \bibinfo {author} {\bibfnamefont
  {M.}~\bibnamefont {Lewenstein}},\ }\href
  {https://doi.org/10.1103/PhysRevLett.83.5198} {\bibfield  {journal} {\bibinfo
   {journal} {Phys. Rev. Lett.}\ }\textbf {\bibinfo {volume} {83}},\ \bibinfo
  {pages} {5198} (\bibinfo {year} {1999})}\BibitemShut {NoStop}%
\bibitem [{\citenamefont {Denschlag}\ \emph {et~al.}(2000)\citenamefont
  {Denschlag}, \citenamefont {Simsarian}, \citenamefont {Feder}, \citenamefont
  {Clark}, \citenamefont {Collins}, \citenamefont {Cubizolles}, \citenamefont
  {Deng}, \citenamefont {Hagley}, \citenamefont {Helmerson}, \citenamefont
  {Reinhardt}, \citenamefont {Rolston}, \citenamefont {Schneider},\ and\
  \citenamefont {Phillips}}]{DENSCHLAG_SCIENCE00}%
  \BibitemOpen
  \bibfield  {author} {\bibinfo {author} {\bibfnamefont {J.}~\bibnamefont
  {Denschlag}}, \bibinfo {author} {\bibfnamefont {J.~E.}\ \bibnamefont
  {Simsarian}}, \bibinfo {author} {\bibfnamefont {D.~L.}\ \bibnamefont
  {Feder}}, \bibinfo {author} {\bibfnamefont {C.~W.}\ \bibnamefont {Clark}},
  \bibinfo {author} {\bibfnamefont {L.~A.}\ \bibnamefont {Collins}}, \bibinfo
  {author} {\bibfnamefont {J.}~\bibnamefont {Cubizolles}}, \bibinfo {author}
  {\bibfnamefont {L.}~\bibnamefont {Deng}}, \bibinfo {author} {\bibfnamefont
  {E.~W.}\ \bibnamefont {Hagley}}, \bibinfo {author} {\bibfnamefont
  {K.}~\bibnamefont {Helmerson}}, \bibinfo {author} {\bibfnamefont {W.~P.}\
  \bibnamefont {Reinhardt}}, \bibinfo {author} {\bibfnamefont {S.~L.}\
  \bibnamefont {Rolston}}, \bibinfo {author} {\bibfnamefont {B.~I.}\
  \bibnamefont {Schneider}},\ and\ \bibinfo {author} {\bibfnamefont {W.~D.}\
  \bibnamefont {Phillips}},\ }\href
  {https://doi.org/10.1126/science.287.5450.97} {\bibfield  {journal} {\bibinfo
   {journal} {Science}\ }\textbf {\bibinfo {volume} {287}},\ \bibinfo {pages}
  {97} (\bibinfo {year} {2000})}\BibitemShut {NoStop}%
\bibitem [{\citenamefont {Anderson}\ \emph {et~al.}(2001)\citenamefont
  {Anderson}, \citenamefont {Haljan}, \citenamefont {Regal}, \citenamefont
  {Feder}, \citenamefont {Collins}, \citenamefont {Clark},\ and\ \citenamefont
  {Cornell}}]{ANDERSON_PRL01}%
  \BibitemOpen
  \bibfield  {author} {\bibinfo {author} {\bibfnamefont {B.~P.}\ \bibnamefont
  {Anderson}}, \bibinfo {author} {\bibfnamefont {P.~C.}\ \bibnamefont
  {Haljan}}, \bibinfo {author} {\bibfnamefont {C.~A.}\ \bibnamefont {Regal}},
  \bibinfo {author} {\bibfnamefont {D.~L.}\ \bibnamefont {Feder}}, \bibinfo
  {author} {\bibfnamefont {L.~A.}\ \bibnamefont {Collins}}, \bibinfo {author}
  {\bibfnamefont {C.~W.}\ \bibnamefont {Clark}},\ and\ \bibinfo {author}
  {\bibfnamefont {E.~A.}\ \bibnamefont {Cornell}},\ }\href
  {https://doi.org/10.1103/PhysRevLett.86.2926} {\bibfield  {journal} {\bibinfo
   {journal} {Phys. Rev. Lett.}\ }\textbf {\bibinfo {volume} {86}},\ \bibinfo
  {pages} {2926} (\bibinfo {year} {2001})}\BibitemShut {NoStop}%
\bibitem [{\citenamefont {Matthews}\ \emph {et~al.}(1999)\citenamefont
  {Matthews}, \citenamefont {Anderson}, \citenamefont {Haljan}, \citenamefont
  {Hall}, \citenamefont {Wieman},\ and\ \citenamefont
  {Cornell}}]{MATTHEWS_PRL99}%
  \BibitemOpen
  \bibfield  {author} {\bibinfo {author} {\bibfnamefont {M.~R.}\ \bibnamefont
  {Matthews}}, \bibinfo {author} {\bibfnamefont {B.~P.}\ \bibnamefont
  {Anderson}}, \bibinfo {author} {\bibfnamefont {P.~C.}\ \bibnamefont
  {Haljan}}, \bibinfo {author} {\bibfnamefont {D.~S.}\ \bibnamefont {Hall}},
  \bibinfo {author} {\bibfnamefont {C.~E.}\ \bibnamefont {Wieman}},\ and\
  \bibinfo {author} {\bibfnamefont {E.~A.}\ \bibnamefont {Cornell}},\ }\href
  {https://doi.org/10.1103/PhysRevLett.83.2498} {\bibfield  {journal} {\bibinfo
   {journal} {Phys. Rev. Lett.}\ }\textbf {\bibinfo {volume} {83}},\ \bibinfo
  {pages} {2498} (\bibinfo {year} {1999})}\BibitemShut {NoStop}%
\bibitem [{\citenamefont {Madison}\ \emph {et~al.}(2000)\citenamefont
  {Madison}, \citenamefont {Chevy}, \citenamefont {Wohlleben},\ and\
  \citenamefont {Dalibard}}]{MADISON_PRL00}%
  \BibitemOpen
  \bibfield  {author} {\bibinfo {author} {\bibfnamefont {K.~W.}\ \bibnamefont
  {Madison}}, \bibinfo {author} {\bibfnamefont {F.}~\bibnamefont {Chevy}},
  \bibinfo {author} {\bibfnamefont {W.}~\bibnamefont {Wohlleben}},\ and\
  \bibinfo {author} {\bibfnamefont {J.}~\bibnamefont {Dalibard}},\ }\href
  {https://doi.org/10.1103/PhysRevLett.84.806} {\bibfield  {journal} {\bibinfo
  {journal} {Phys. Rev. Lett.}\ }\textbf {\bibinfo {volume} {84}},\ \bibinfo
  {pages} {806} (\bibinfo {year} {2000})}\BibitemShut {NoStop}%
\bibitem [{\citenamefont {Billy}\ \emph {et~al.}(2008)\citenamefont {Billy},
  \citenamefont {Josse}, \citenamefont {Zuo}, \citenamefont {Bernard},
  \citenamefont {Hambrecht}, \citenamefont {Lugan}, \citenamefont
  {Cl{\'{e}}ment}, \citenamefont {Sanchez-Palencia}, \citenamefont {Bouyer},\
  and\ \citenamefont {Aspect}}]{BILLY_NATURE08}%
  \BibitemOpen
  \bibfield  {author} {\bibinfo {author} {\bibfnamefont {J.}~\bibnamefont
  {Billy}}, \bibinfo {author} {\bibfnamefont {V.}~\bibnamefont {Josse}},
  \bibinfo {author} {\bibfnamefont {Z.}~\bibnamefont {Zuo}}, \bibinfo {author}
  {\bibfnamefont {A.}~\bibnamefont {Bernard}}, \bibinfo {author} {\bibfnamefont
  {B.}~\bibnamefont {Hambrecht}}, \bibinfo {author} {\bibfnamefont
  {P.}~\bibnamefont {Lugan}}, \bibinfo {author} {\bibfnamefont
  {D.}~\bibnamefont {Cl{\'{e}}ment}}, \bibinfo {author} {\bibfnamefont
  {L.}~\bibnamefont {Sanchez-Palencia}}, \bibinfo {author} {\bibfnamefont
  {P.}~\bibnamefont {Bouyer}},\ and\ \bibinfo {author} {\bibfnamefont
  {A.}~\bibnamefont {Aspect}},\ }\href {https://doi.org/10.1038/nature07000}
  {\bibfield  {journal} {\bibinfo  {journal} {Nature}\ }\textbf {\bibinfo
  {volume} {453}},\ \bibinfo {pages} {891} (\bibinfo {year}
  {2008})}\BibitemShut {NoStop}%
\bibitem [{\citenamefont {Roati}\ \emph {et~al.}(2008)\citenamefont {Roati},
  \citenamefont {D'Errico}, \citenamefont {Fallani}, \citenamefont {Fattori},
  \citenamefont {Fort}, \citenamefont {Zaccanti}, \citenamefont {Modugno},
  \citenamefont {Modugno},\ and\ \citenamefont {Inguscio}}]{ROATI_NATURE08}%
  \BibitemOpen
  \bibfield  {author} {\bibinfo {author} {\bibfnamefont {G.}~\bibnamefont
  {Roati}}, \bibinfo {author} {\bibfnamefont {C.}~\bibnamefont {D'Errico}},
  \bibinfo {author} {\bibfnamefont {L.}~\bibnamefont {Fallani}}, \bibinfo
  {author} {\bibfnamefont {M.}~\bibnamefont {Fattori}}, \bibinfo {author}
  {\bibfnamefont {C.}~\bibnamefont {Fort}}, \bibinfo {author} {\bibfnamefont
  {M.}~\bibnamefont {Zaccanti}}, \bibinfo {author} {\bibfnamefont
  {G.}~\bibnamefont {Modugno}}, \bibinfo {author} {\bibfnamefont
  {M.}~\bibnamefont {Modugno}},\ and\ \bibinfo {author} {\bibfnamefont
  {M.}~\bibnamefont {Inguscio}},\ }\href {https://doi.org/10.1038/nature07071}
  {\bibfield  {journal} {\bibinfo  {journal} {Nature}\ }\textbf {\bibinfo
  {volume} {453}},\ \bibinfo {pages} {895} (\bibinfo {year}
  {2008})}\BibitemShut {NoStop}%
\bibitem [{\citenamefont {{Di Carli}}\ \emph {et~al.}(2019)\citenamefont {{Di
  Carli}}, \citenamefont {Colquhoun}, \citenamefont {Henderson}, \citenamefont
  {Flannigan}, \citenamefont {Oppo}, \citenamefont {Daley}, \citenamefont
  {Kuhr},\ and\ \citenamefont {Haller}}]{DICARLI_PRL19}%
  \BibitemOpen
  \bibfield  {author} {\bibinfo {author} {\bibfnamefont {A.}~\bibnamefont {{Di
  Carli}}}, \bibinfo {author} {\bibfnamefont {C.~D.}\ \bibnamefont
  {Colquhoun}}, \bibinfo {author} {\bibfnamefont {G.}~\bibnamefont
  {Henderson}}, \bibinfo {author} {\bibfnamefont {S.}~\bibnamefont
  {Flannigan}}, \bibinfo {author} {\bibfnamefont {G.-L.}\ \bibnamefont {Oppo}},
  \bibinfo {author} {\bibfnamefont {A.~J.}\ \bibnamefont {Daley}}, \bibinfo
  {author} {\bibfnamefont {S.}~\bibnamefont {Kuhr}},\ and\ \bibinfo {author}
  {\bibfnamefont {E.}~\bibnamefont {Haller}},\ }\href
  {https://doi.org/10.1103/PhysRevLett.123.123602} {\bibfield  {journal}
  {\bibinfo  {journal} {Phys. Rev. Lett.}\ }\textbf {\bibinfo {volume} {123}},\
  \bibinfo {pages} {123602} (\bibinfo {year} {2019})}\BibitemShut {NoStop}%
\bibitem [{\citenamefont {Luo}\ \emph {et~al.}(2020)\citenamefont {Luo},
  \citenamefont {Jin}, \citenamefont {Nguyen}, \citenamefont {Malomed},
  \citenamefont {Marchukov}, \citenamefont {Yurovsky}, \citenamefont {Dunjko},
  \citenamefont {Olshanii},\ and\ \citenamefont {Hulet}}]{LUO_PRL20}%
  \BibitemOpen
  \bibfield  {author} {\bibinfo {author} {\bibfnamefont {D.}~\bibnamefont
  {Luo}}, \bibinfo {author} {\bibfnamefont {Y.}~\bibnamefont {Jin}}, \bibinfo
  {author} {\bibfnamefont {J.~H.~V.}\ \bibnamefont {Nguyen}}, \bibinfo {author}
  {\bibfnamefont {B.~A.}\ \bibnamefont {Malomed}}, \bibinfo {author}
  {\bibfnamefont {O.~V.}\ \bibnamefont {Marchukov}}, \bibinfo {author}
  {\bibfnamefont {V.~A.}\ \bibnamefont {Yurovsky}}, \bibinfo {author}
  {\bibfnamefont {V.}~\bibnamefont {Dunjko}}, \bibinfo {author} {\bibfnamefont
  {M.}~\bibnamefont {Olshanii}},\ and\ \bibinfo {author} {\bibfnamefont
  {R.~G.}\ \bibnamefont {Hulet}},\ }\href
  {https://doi.org/10.1103/PhysRevLett.125.183902} {\bibfield  {journal}
  {\bibinfo  {journal} {Phys. Rev. Lett.}\ }\textbf {\bibinfo {volume} {125}},\
  \bibinfo {pages} {183902} (\bibinfo {year} {2020})}\BibitemShut {NoStop}%
\bibitem [{\citenamefont {Cabrera}\ \emph {et~al.}(2018)\citenamefont
  {Cabrera}, \citenamefont {Tanzi}, \citenamefont {Sanz}, \citenamefont
  {Naylor}, \citenamefont {Thomas}, \citenamefont {Cheiney},\ and\
  \citenamefont {Tarruell}}]{CABRERA_SCIENCE18}%
  \BibitemOpen
  \bibfield  {author} {\bibinfo {author} {\bibfnamefont {C.~R.}\ \bibnamefont
  {Cabrera}}, \bibinfo {author} {\bibfnamefont {L.}~\bibnamefont {Tanzi}},
  \bibinfo {author} {\bibfnamefont {J.}~\bibnamefont {Sanz}}, \bibinfo {author}
  {\bibfnamefont {B.}~\bibnamefont {Naylor}}, \bibinfo {author} {\bibfnamefont
  {P.}~\bibnamefont {Thomas}}, \bibinfo {author} {\bibfnamefont
  {P.}~\bibnamefont {Cheiney}},\ and\ \bibinfo {author} {\bibfnamefont
  {L.}~\bibnamefont {Tarruell}},\ }\href
  {https://doi.org/10.1126/science.aao5686} {\bibfield  {journal} {\bibinfo
  {journal} {Science}\ }\textbf {\bibinfo {volume} {359}},\ \bibinfo {pages}
  {301} (\bibinfo {year} {2018})}\BibitemShut {NoStop}%
\bibitem [{\citenamefont {Cheiney}\ \emph {et~al.}(2018)\citenamefont
  {Cheiney}, \citenamefont {Cabrera}, \citenamefont {Sanz}, \citenamefont
  {Naylor}, \citenamefont {Tanzi},\ and\ \citenamefont
  {Tarruell}}]{CHEINEY_PRL18}%
  \BibitemOpen
  \bibfield  {author} {\bibinfo {author} {\bibfnamefont {P.}~\bibnamefont
  {Cheiney}}, \bibinfo {author} {\bibfnamefont {C.~R.}\ \bibnamefont
  {Cabrera}}, \bibinfo {author} {\bibfnamefont {J.}~\bibnamefont {Sanz}},
  \bibinfo {author} {\bibfnamefont {B.}~\bibnamefont {Naylor}}, \bibinfo
  {author} {\bibfnamefont {L.}~\bibnamefont {Tanzi}},\ and\ \bibinfo {author}
  {\bibfnamefont {L.}~\bibnamefont {Tarruell}},\ }\href
  {https://doi.org/10.1103/PhysRevLett.120.135301} {\bibfield  {journal}
  {\bibinfo  {journal} {Phys. Rev. Lett.}\ }\textbf {\bibinfo {volume} {120}},\
  \bibinfo {pages} {135301} (\bibinfo {year} {2018})}\BibitemShut {NoStop}%
\bibitem [{\citenamefont {Semeghini}\ \emph {et~al.}(2018)\citenamefont
  {Semeghini}, \citenamefont {Ferioli}, \citenamefont {Masi}, \citenamefont
  {Mazzinghi}, \citenamefont {Wolswijk}, \citenamefont {Minardi}, \citenamefont
  {Modugno}, \citenamefont {Modugno}, \citenamefont {Inguscio},\ and\
  \citenamefont {Fattori}}]{SEMEGHINI_PRL18}%
  \BibitemOpen
  \bibfield  {author} {\bibinfo {author} {\bibfnamefont {G.}~\bibnamefont
  {Semeghini}}, \bibinfo {author} {\bibfnamefont {G.}~\bibnamefont {Ferioli}},
  \bibinfo {author} {\bibfnamefont {L.}~\bibnamefont {Masi}}, \bibinfo {author}
  {\bibfnamefont {C.}~\bibnamefont {Mazzinghi}}, \bibinfo {author}
  {\bibfnamefont {L.}~\bibnamefont {Wolswijk}}, \bibinfo {author}
  {\bibfnamefont {F.}~\bibnamefont {Minardi}}, \bibinfo {author} {\bibfnamefont
  {M.}~\bibnamefont {Modugno}}, \bibinfo {author} {\bibfnamefont
  {G.}~\bibnamefont {Modugno}}, \bibinfo {author} {\bibfnamefont
  {M.}~\bibnamefont {Inguscio}},\ and\ \bibinfo {author} {\bibfnamefont
  {M.}~\bibnamefont {Fattori}},\ }\href
  {https://doi.org/10.1103/PhysRevLett.120.235301} {\bibfield  {journal}
  {\bibinfo  {journal} {Phys. Rev. Lett.}\ }\textbf {\bibinfo {volume} {120}},\
  \bibinfo {pages} {235301} (\bibinfo {year} {2018})}\BibitemShut {NoStop}%
\bibitem [{\citenamefont {D'Errico}\ \emph {et~al.}(2019)\citenamefont
  {D'Errico}, \citenamefont {Burchianti}, \citenamefont {Prevedelli},
  \citenamefont {Salasnich}, \citenamefont {Ancilotto}, \citenamefont
  {Modugno}, \citenamefont {Minardi},\ and\ \citenamefont
  {Fort}}]{DERRICO_PRR19}%
  \BibitemOpen
  \bibfield  {author} {\bibinfo {author} {\bibfnamefont {C.}~\bibnamefont
  {D'Errico}}, \bibinfo {author} {\bibfnamefont {A.}~\bibnamefont
  {Burchianti}}, \bibinfo {author} {\bibfnamefont {M.}~\bibnamefont
  {Prevedelli}}, \bibinfo {author} {\bibfnamefont {L.}~\bibnamefont
  {Salasnich}}, \bibinfo {author} {\bibfnamefont {F.}~\bibnamefont
  {Ancilotto}}, \bibinfo {author} {\bibfnamefont {M.}~\bibnamefont {Modugno}},
  \bibinfo {author} {\bibfnamefont {F.}~\bibnamefont {Minardi}},\ and\ \bibinfo
  {author} {\bibfnamefont {C.}~\bibnamefont {Fort}},\ }\href
  {https://doi.org/10.1103/PhysRevResearch.1.033155} {\bibfield  {journal}
  {\bibinfo  {journal} {Phys. Rev. Res.}\ }\textbf {\bibinfo {volume} {1}},\
  \bibinfo {pages} {033155} (\bibinfo {year} {2019})}\BibitemShut {NoStop}%
\bibitem [{\citenamefont {Jain}\ and\ \citenamefont
  {Boninsegni}(2011)}]{JAIN_PRA11}%
  \BibitemOpen
  \bibfield  {author} {\bibinfo {author} {\bibfnamefont {P.}~\bibnamefont
  {Jain}}\ and\ \bibinfo {author} {\bibfnamefont {M.}~\bibnamefont
  {Boninsegni}},\ }\href {https://doi.org/10.1103/PhysRevA.83.023602}
  {\bibfield  {journal} {\bibinfo  {journal} {Phys. Rev. A}\ }\textbf {\bibinfo
  {volume} {83}},\ \bibinfo {pages} {023602} (\bibinfo {year}
  {2011})}\BibitemShut {NoStop}%
\bibitem [{\citenamefont {dos Santos}\ and\ \citenamefont
  {Cardoso}(2021)}]{dosSANTOS_PRE21}%
  \BibitemOpen
  \bibfield  {author} {\bibinfo {author} {\bibfnamefont {M.~C.~P.}\
  \bibnamefont {dos Santos}}\ and\ \bibinfo {author} {\bibfnamefont {W.~B.}\
  \bibnamefont {Cardoso}},\ }\href
  {https://doi.org/10.1103/PhysRevE.103.052210} {\bibfield  {journal} {\bibinfo
   {journal} {Phys. Rev. E}\ }\textbf {\bibinfo {volume} {103}},\ \bibinfo
  {pages} {052210} (\bibinfo {year} {2021})}\BibitemShut {NoStop}%
\bibitem [{\citenamefont {Xu}\ \emph {et~al.}(2008)\citenamefont {Xu},
  \citenamefont {Lu},\ and\ \citenamefont {Li}}]{XU_PRA08}%
  \BibitemOpen
  \bibfield  {author} {\bibinfo {author} {\bibfnamefont {X.-Q.}\ \bibnamefont
  {Xu}}, \bibinfo {author} {\bibfnamefont {L.-H.}\ \bibnamefont {Lu}},\ and\
  \bibinfo {author} {\bibfnamefont {Y.-Q.}\ \bibnamefont {Li}},\ }\href
  {https://doi.org/10.1103/PhysRevA.78.043609} {\bibfield  {journal} {\bibinfo
  {journal} {Phys. Rev. A}\ }\textbf {\bibinfo {volume} {78}},\ \bibinfo
  {pages} {043609} (\bibinfo {year} {2008})}\BibitemShut {NoStop}%
\bibitem [{\citenamefont {Satija}\ \emph {et~al.}(2009)\citenamefont {Satija},
  \citenamefont {Balakrishnan}, \citenamefont {Naudus}, \citenamefont {Heward},
  \citenamefont {Edwards},\ and\ \citenamefont {Clark}}]{SATIJA_PRA09}%
  \BibitemOpen
  \bibfield  {author} {\bibinfo {author} {\bibfnamefont {I.~I.}\ \bibnamefont
  {Satija}}, \bibinfo {author} {\bibfnamefont {R.}~\bibnamefont
  {Balakrishnan}}, \bibinfo {author} {\bibfnamefont {P.}~\bibnamefont
  {Naudus}}, \bibinfo {author} {\bibfnamefont {J.}~\bibnamefont {Heward}},
  \bibinfo {author} {\bibfnamefont {M.}~\bibnamefont {Edwards}},\ and\ \bibinfo
  {author} {\bibfnamefont {C.~W.}\ \bibnamefont {Clark}},\ }\href
  {https://doi.org/10.1103/PhysRevA.79.033616} {\bibfield  {journal} {\bibinfo
  {journal} {Phys. Rev. A}\ }\textbf {\bibinfo {volume} {79}},\ \bibinfo
  {pages} {033616} (\bibinfo {year} {2009})}\BibitemShut {NoStop}%
\bibitem [{\citenamefont {Mazzarella}\ \emph {et~al.}(2009)\citenamefont
  {Mazzarella}, \citenamefont {Moratti}, \citenamefont {Salasnich},
  \citenamefont {Salerno},\ and\ \citenamefont {Toigo}}]{MAZZARELLA_JPB09}%
  \BibitemOpen
  \bibfield  {author} {\bibinfo {author} {\bibfnamefont {G.}~\bibnamefont
  {Mazzarella}}, \bibinfo {author} {\bibfnamefont {M.}~\bibnamefont {Moratti}},
  \bibinfo {author} {\bibfnamefont {L.}~\bibnamefont {Salasnich}}, \bibinfo
  {author} {\bibfnamefont {M.}~\bibnamefont {Salerno}},\ and\ \bibinfo {author}
  {\bibfnamefont {F.}~\bibnamefont {Toigo}},\ }\href
  {https://doi.org/10.1088/0953-4075/42/12/125301} {\bibfield  {journal}
  {\bibinfo  {journal} {J. Phys. B At. Mol. Opt. Phys.}\ }\textbf {\bibinfo
  {volume} {42}},\ \bibinfo {pages} {125301} (\bibinfo {year}
  {2009})}\BibitemShut {NoStop}%
\bibitem [{\citenamefont {Acus}\ \emph {et~al.}(2012)\citenamefont {Acus},
  \citenamefont {Malomed},\ and\ \citenamefont {Shnir}}]{ACUS_PD12}%
  \BibitemOpen
  \bibfield  {author} {\bibinfo {author} {\bibfnamefont {A.}~\bibnamefont
  {Acus}}, \bibinfo {author} {\bibfnamefont {B.~A.}\ \bibnamefont {Malomed}},\
  and\ \bibinfo {author} {\bibfnamefont {Y.}~\bibnamefont {Shnir}},\ }\href
  {https://doi.org/10.1016/j.physd.2012.02.012} {\bibfield  {journal} {\bibinfo
   {journal} {Phys. D Nonlinear Phenom.}\ }\textbf {\bibinfo {volume} {241}},\
  \bibinfo {pages} {987} (\bibinfo {year} {2012})}\BibitemShut {NoStop}%
\bibitem [{\citenamefont {Adhikari}\ \emph {et~al.}(2010)\citenamefont
  {Adhikari}, \citenamefont {Malomed}, \citenamefont {Salasnich},\ and\
  \citenamefont {Toigo}}]{ADHIKARI_PRA10}%
  \BibitemOpen
  \bibfield  {author} {\bibinfo {author} {\bibfnamefont {S.~K.}\ \bibnamefont
  {Adhikari}}, \bibinfo {author} {\bibfnamefont {B.~A.}\ \bibnamefont
  {Malomed}}, \bibinfo {author} {\bibfnamefont {L.}~\bibnamefont {Salasnich}},\
  and\ \bibinfo {author} {\bibfnamefont {F.}~\bibnamefont {Toigo}},\ }\href
  {https://doi.org/10.1103/PhysRevA.81.053630} {\bibfield  {journal} {\bibinfo
  {journal} {Phys. Rev. A}\ }\textbf {\bibinfo {volume} {81}},\ \bibinfo
  {pages} {053630} (\bibinfo {year} {2010})}\BibitemShut {NoStop}%
\bibitem [{\citenamefont {Hacker}\ and\ \citenamefont
  {Malomed}(2021)}]{HACKER_SYMMETRY21}%
  \BibitemOpen
  \bibfield  {author} {\bibinfo {author} {\bibfnamefont {N.}~\bibnamefont
  {Hacker}}\ and\ \bibinfo {author} {\bibfnamefont {B.~A.}\ \bibnamefont
  {Malomed}},\ }\href {https://doi.org/10.3390/sym13030372} {\bibfield
  {journal} {\bibinfo  {journal} {Symmetry (Basel).}\ }\textbf {\bibinfo
  {volume} {13}},\ \bibinfo {pages} {372} (\bibinfo {year} {2021})}\BibitemShut
  {NoStop}%
\bibitem [{\citenamefont {Espinosa-Cer{\'{o}}n}\ \emph
  {et~al.}(2012)\citenamefont {Espinosa-Cer{\'{o}}n}, \citenamefont {Malomed},
  \citenamefont {Fujioka},\ and\ \citenamefont
  {Rodr{\'{i}}guez}}]{ESPINOSA-CERON_CHAOS12}%
  \BibitemOpen
  \bibfield  {author} {\bibinfo {author} {\bibfnamefont {A.}~\bibnamefont
  {Espinosa-Cer{\'{o}}n}}, \bibinfo {author} {\bibfnamefont {B.~A.}\
  \bibnamefont {Malomed}}, \bibinfo {author} {\bibfnamefont {J.}~\bibnamefont
  {Fujioka}},\ and\ \bibinfo {author} {\bibfnamefont {R.~F.}\ \bibnamefont
  {Rodr{\'{i}}guez}},\ }\href {https://doi.org/10.1063/1.4752244} {\bibfield
  {journal} {\bibinfo  {journal} {Chaos An Interdiscip. J. Nonlinear Sci.}\
  }\textbf {\bibinfo {volume} {22}},\ \bibinfo {pages} {033145} (\bibinfo
  {year} {2012})}\BibitemShut {NoStop}%
\bibitem [{\citenamefont {Pitaevskii}\ and\ \citenamefont
  {Stringari}(2003)}]{Pitaevskii_03}%
  \BibitemOpen
  \bibfield  {author} {\bibinfo {author} {\bibfnamefont {L.~P.}\ \bibnamefont
  {Pitaevskii}}\ and\ \bibinfo {author} {\bibfnamefont {S.}~\bibnamefont
  {Stringari}},\ }\href {https://books.google.com.br/books?id=rIobbOxC4j4C}
  {\emph {\bibinfo {title} {{Bose--Einstein Condensation}}}},\ International
  Series of Monographs on Physics\ (\bibinfo  {publisher} {Clarendon Press},\
  \bibinfo {year} {2003})\BibitemShut {NoStop}%
\bibitem [{\citenamefont {Pethick}\ and\ \citenamefont
  {Smith}(2008)}]{PETHICK_08}%
  \BibitemOpen
  \bibfield  {author} {\bibinfo {author} {\bibfnamefont {C.~J.}\ \bibnamefont
  {Pethick}}\ and\ \bibinfo {author} {\bibfnamefont {H.}~\bibnamefont
  {Smith}},\ }\href {https://doi.org/10.1017/CBO9780511802850} {\emph {\bibinfo
  {title} {{Bose--Einstein Condensation in Dilute Gases}}}}\ (\bibinfo
  {publisher} {Cambridge University Press},\ \bibinfo {address} {Cambridge},\
  \bibinfo {year} {2008})\BibitemShut {NoStop}%
\bibitem [{\citenamefont {Salasnich}\ \emph
  {et~al.}(2002{\natexlab{a}})\citenamefont {Salasnich}, \citenamefont
  {Parola},\ and\ \citenamefont {Reatto}}]{SALASNICH_PRA02}%
  \BibitemOpen
  \bibfield  {author} {\bibinfo {author} {\bibfnamefont {L.}~\bibnamefont
  {Salasnich}}, \bibinfo {author} {\bibfnamefont {A.}~\bibnamefont {Parola}},\
  and\ \bibinfo {author} {\bibfnamefont {L.}~\bibnamefont {Reatto}},\ }\href
  {https://doi.org/10.1103/PhysRevA.65.043614} {\bibfield  {journal} {\bibinfo
  {journal} {Phys. Rev. A}\ }\textbf {\bibinfo {volume} {65}},\ \bibinfo
  {pages} {043614} (\bibinfo {year} {2002}{\natexlab{a}})}\BibitemShut
  {NoStop}%
\bibitem [{\citenamefont {Massignan}\ and\ \citenamefont
  {Modugno}(2003)}]{MASSIGNAN_PRA03}%
  \BibitemOpen
  \bibfield  {author} {\bibinfo {author} {\bibfnamefont {P.}~\bibnamefont
  {Massignan}}\ and\ \bibinfo {author} {\bibfnamefont {M.}~\bibnamefont
  {Modugno}},\ }\href {https://doi.org/10.1103/PhysRevA.67.023614} {\bibfield
  {journal} {\bibinfo  {journal} {Phys. Rev. A}\ }\textbf {\bibinfo {volume}
  {67}},\ \bibinfo {pages} {023614} (\bibinfo {year} {2003})}\BibitemShut
  {NoStop}%
\bibitem [{\citenamefont {Buitrago}\ and\ \citenamefont
  {Adhikari}(2009)}]{BUITRAGO_JPB09}%
  \BibitemOpen
  \bibfield  {author} {\bibinfo {author} {\bibfnamefont {C.~A.~G.}\
  \bibnamefont {Buitrago}}\ and\ \bibinfo {author} {\bibfnamefont {S.~K.}\
  \bibnamefont {Adhikari}},\ }\href
  {https://doi.org/10.1088/0953-4075/42/21/215306} {\bibfield  {journal}
  {\bibinfo  {journal} {J. Phys. B At. Mol. Opt. Phys.}\ }\textbf {\bibinfo
  {volume} {42}},\ \bibinfo {pages} {215306} (\bibinfo {year}
  {2009})}\BibitemShut {NoStop}%
\bibitem [{\citenamefont {Mateo}\ and\ \citenamefont
  {Delgado}(2008)}]{Mateo_PRA08}%
  \BibitemOpen
  \bibfield  {author} {\bibinfo {author} {\bibfnamefont {A.~M.}\ \bibnamefont
  {Mateo}}\ and\ \bibinfo {author} {\bibfnamefont {V.}~\bibnamefont
  {Delgado}},\ }\href {https://doi.org/10.1103/PhysRevA.77.013617} {\bibfield
  {journal} {\bibinfo  {journal} {Phys. Rev. A}\ }\textbf {\bibinfo {volume}
  {77}},\ \bibinfo {pages} {013617} (\bibinfo {year} {2008})}\BibitemShut
  {NoStop}%
\bibitem [{\citenamefont {Couto}\ \emph {et~al.}(2018)\citenamefont {Couto},
  \citenamefont {Avelar},\ and\ \citenamefont {Cardoso}}]{Couto_AP18}%
  \BibitemOpen
  \bibfield  {author} {\bibinfo {author} {\bibfnamefont {H.~L.~C.}\
  \bibnamefont {Couto}}, \bibinfo {author} {\bibfnamefont {A.~T.}\ \bibnamefont
  {Avelar}},\ and\ \bibinfo {author} {\bibfnamefont {W.~B.}\ \bibnamefont
  {Cardoso}},\ }\href {https://doi.org/10.1002/andp.201700352} {\bibfield
  {journal} {\bibinfo  {journal} {Ann. Phys.}\ }\textbf {\bibinfo {volume}
  {530}},\ \bibinfo {pages} {1700352} (\bibinfo {year} {2018})}\BibitemShut
  {NoStop}%
\bibitem [{\citenamefont {Salasnich}\ \emph
  {et~al.}(2002{\natexlab{b}})\citenamefont {Salasnich}, \citenamefont
  {Parola},\ and\ \citenamefont {Reatto}}]{SALASNICH_PRA02_2}%
  \BibitemOpen
  \bibfield  {author} {\bibinfo {author} {\bibfnamefont {L.}~\bibnamefont
  {Salasnich}}, \bibinfo {author} {\bibfnamefont {A.}~\bibnamefont {Parola}},\
  and\ \bibinfo {author} {\bibfnamefont {L.}~\bibnamefont {Reatto}},\ }\href
  {https://doi.org/10.1103/PhysRevA.66.043603} {\bibfield  {journal} {\bibinfo
  {journal} {Phys. Rev. A}\ }\textbf {\bibinfo {volume} {66}},\ \bibinfo
  {pages} {043603} (\bibinfo {year} {2002}{\natexlab{b}})}\BibitemShut
  {NoStop}%
\bibitem [{\citenamefont {Salasnich}\ and\ \citenamefont
  {Malomed}(2006)}]{SALASNICH_PRA06}%
  \BibitemOpen
  \bibfield  {author} {\bibinfo {author} {\bibfnamefont {L.}~\bibnamefont
  {Salasnich}}\ and\ \bibinfo {author} {\bibfnamefont {B.~A.}\ \bibnamefont
  {Malomed}},\ }\href {https://doi.org/10.1103/PhysRevA.74.053610} {\bibfield
  {journal} {\bibinfo  {journal} {Phys. Rev. A}\ }\textbf {\bibinfo {volume}
  {74}},\ \bibinfo {pages} {053610} (\bibinfo {year} {2006})}\BibitemShut
  {NoStop}%
\bibitem [{\citenamefont {Salasnich}\ \emph {et~al.}(2007)\citenamefont
  {Salasnich}, \citenamefont {Cetoli}, \citenamefont {Malomed}, \citenamefont
  {Toigo},\ and\ \citenamefont {Reatto}}]{SALASNICH_PRA07}%
  \BibitemOpen
  \bibfield  {author} {\bibinfo {author} {\bibfnamefont {L.}~\bibnamefont
  {Salasnich}}, \bibinfo {author} {\bibfnamefont {A.}~\bibnamefont {Cetoli}},
  \bibinfo {author} {\bibfnamefont {B.~A.}\ \bibnamefont {Malomed}}, \bibinfo
  {author} {\bibfnamefont {F.}~\bibnamefont {Toigo}},\ and\ \bibinfo {author}
  {\bibfnamefont {L.}~\bibnamefont {Reatto}},\ }\href
  {https://doi.org/10.1103/PhysRevA.76.013623} {\bibfield  {journal} {\bibinfo
  {journal} {Phys. Rev. A}\ }\textbf {\bibinfo {volume} {76}},\ \bibinfo
  {pages} {013623} (\bibinfo {year} {2007})}\BibitemShut {NoStop}%
\bibitem [{\citenamefont {Adhikari}\ and\ \citenamefont
  {Salasnich}(2009)}]{ADHIKARI_NJP09}%
  \BibitemOpen
  \bibfield  {author} {\bibinfo {author} {\bibfnamefont {S.~K.}\ \bibnamefont
  {Adhikari}}\ and\ \bibinfo {author} {\bibfnamefont {L.}~\bibnamefont
  {Salasnich}},\ }\href {https://doi.org/10.1088/1367-2630/11/2/023011}
  {\bibfield  {journal} {\bibinfo  {journal} {New J. Phys.}\ }\textbf {\bibinfo
  {volume} {11}},\ \bibinfo {pages} {023011} (\bibinfo {year}
  {2009})}\BibitemShut {NoStop}%
\bibitem [{\citenamefont {Cardoso}\ \emph {et~al.}(2011)\citenamefont
  {Cardoso}, \citenamefont {Avelar},\ and\ \citenamefont
  {Bazeia}}]{Cardoso_PRE11}%
  \BibitemOpen
  \bibfield  {author} {\bibinfo {author} {\bibfnamefont {W.~B.}\ \bibnamefont
  {Cardoso}}, \bibinfo {author} {\bibfnamefont {A.~T.}\ \bibnamefont
  {Avelar}},\ and\ \bibinfo {author} {\bibfnamefont {D.}~\bibnamefont
  {Bazeia}},\ }\href {https://doi.org/10.1103/PhysRevE.83.036604} {\bibfield
  {journal} {\bibinfo  {journal} {Phys. Rev. E}\ }\textbf {\bibinfo {volume}
  {83}},\ \bibinfo {pages} {36604} (\bibinfo {year} {2011})}\BibitemShut
  {NoStop}%
\bibitem [{\citenamefont {dos Santos}\ and\ \citenamefont
  {Cardoso}(2019)}]{dosSantos_PLA19}%
  \BibitemOpen
  \bibfield  {author} {\bibinfo {author} {\bibfnamefont {M.~C.}\ \bibnamefont
  {dos Santos}}\ and\ \bibinfo {author} {\bibfnamefont {W.~B.}\ \bibnamefont
  {Cardoso}},\ }\href {https://doi.org/10.1016/j.physleta.2019.01.064}
  {\bibfield  {journal} {\bibinfo  {journal} {Phys. Lett. A}\ }\textbf
  {\bibinfo {volume} {383}},\ \bibinfo {pages} {1435} (\bibinfo {year}
  {2019})}\BibitemShut {NoStop}%
\bibitem [{\citenamefont {dos Santos}\ \emph {et~al.}(2019)\citenamefont {dos
  Santos}, \citenamefont {Malomed},\ and\ \citenamefont
  {Cardoso}}]{Santos_JPB19}%
  \BibitemOpen
  \bibfield  {author} {\bibinfo {author} {\bibfnamefont {M.~C.~P.}\
  \bibnamefont {dos Santos}}, \bibinfo {author} {\bibfnamefont {B.~A.}\
  \bibnamefont {Malomed}},\ and\ \bibinfo {author} {\bibfnamefont {W.~B.}\
  \bibnamefont {Cardoso}},\ }\href {https://doi.org/10.1088/1361-6455/ab4fb7}
  {\bibfield  {journal} {\bibinfo  {journal} {J. Phys. B At. Mol. Opt. Phys.}\
  }\textbf {\bibinfo {volume} {52}},\ \bibinfo {pages} {245301} (\bibinfo
  {year} {2019})}\BibitemShut {NoStop}%
\bibitem [{\citenamefont {dos Santos}\ \emph {et~al.}(2021)\citenamefont {dos
  Santos}, \citenamefont {Cardoso},\ and\ \citenamefont
  {Malomed}}]{dosSANTOS_EPJST21}%
  \BibitemOpen
  \bibfield  {author} {\bibinfo {author} {\bibfnamefont {M.~C.~P.}\
  \bibnamefont {dos Santos}}, \bibinfo {author} {\bibfnamefont {W.~B.}\
  \bibnamefont {Cardoso}},\ and\ \bibinfo {author} {\bibfnamefont {B.~A.}\
  \bibnamefont {Malomed}},\ }\bibfield  {journal} {\bibinfo  {journal} {Eur.
  Phys. J. Spec. Top.}\ }\href
  {https://doi.org/10.1140/epjs/s11734-021-00351-2}
  {10.1140/epjs/s11734-021-00351-2} (\bibinfo {year} {2021})\BibitemShut
  {NoStop}%
\bibitem [{\citenamefont {Young-S.}\ \emph {et~al.}(2010)\citenamefont
  {Young-S.}, \citenamefont {Salasnich},\ and\ \citenamefont
  {Adhikari}}]{Young_PRA10}%
  \BibitemOpen
  \bibfield  {author} {\bibinfo {author} {\bibfnamefont {L.~E.}\ \bibnamefont
  {Young-S.}}, \bibinfo {author} {\bibfnamefont {L.}~\bibnamefont
  {Salasnich}},\ and\ \bibinfo {author} {\bibfnamefont {S.~K.}\ \bibnamefont
  {Adhikari}},\ }\href {https://doi.org/10.1103/PhysRevA.82.053601} {\bibfield
  {journal} {\bibinfo  {journal} {Phys. Rev. A}\ }\textbf {\bibinfo {volume}
  {82}},\ \bibinfo {pages} {053601} (\bibinfo {year} {2010})}\BibitemShut
  {NoStop}%
\bibitem [{\citenamefont {Adhikari}(2011)}]{Adhikari_JPB11}%
  \BibitemOpen
  \bibfield  {author} {\bibinfo {author} {\bibfnamefont {S.~K.}\ \bibnamefont
  {Adhikari}},\ }\href {https://doi.org/10.1088/0953-4075/44/7/075301}
  {\bibfield  {journal} {\bibinfo  {journal} {J. Phys. B At. Mol. Opt. Phys.}\
  }\textbf {\bibinfo {volume} {44}},\ \bibinfo {pages} {075301} (\bibinfo
  {year} {2011})}\BibitemShut {NoStop}%
\bibitem [{\citenamefont {Landau}\ and\ \citenamefont
  {Lifshitz}(1959)}]{landau2013quantum}%
  \BibitemOpen
  \bibfield  {author} {\bibinfo {author} {\bibfnamefont {L.}~\bibnamefont
  {Landau}}\ and\ \bibinfo {author} {\bibfnamefont {L.}~\bibnamefont
  {Lifshitz}},\ }\href@noop {} {\emph {\bibinfo {title} {{Course in Theoretical
  Physics (Quantum Mechanics: Non-Relativistic Theory)}}}},\ Vol.~\bibinfo
  {volume} {3}\ (\bibinfo  {publisher} {New York: Pergamon},\ \bibinfo {year}
  {1959})\BibitemShut {NoStop}%
\bibitem [{\citenamefont {P{\"o}schl}\ and\ \citenamefont
  {Teller}(1933)}]{Posh_ZP33}%
  \BibitemOpen
  \bibfield  {author} {\bibinfo {author} {\bibfnamefont {G.}~\bibnamefont
  {P{\"o}schl}}\ and\ \bibinfo {author} {\bibfnamefont {E.}~\bibnamefont
  {Teller}},\ }\href {https://doi.org/10.1007/BF01331132} {\bibfield  {journal}
  {\bibinfo  {journal} {Zeitschrift f{\"u}r Phys.}\ }\textbf {\bibinfo {volume}
  {83}},\ \bibinfo {pages} {143} (\bibinfo {year} {1933})}\BibitemShut
  {NoStop}%
\bibitem [{\citenamefont {Kiriushcheva}\ and\ \citenamefont
  {Kuzmin}(1998)}]{Kiriushcheva_AJP98}%
  \BibitemOpen
  \bibfield  {author} {\bibinfo {author} {\bibfnamefont {N.}~\bibnamefont
  {Kiriushcheva}}\ and\ \bibinfo {author} {\bibfnamefont {S.}~\bibnamefont
  {Kuzmin}},\ }\href {https://doi.org/10.1119/1.18985} {\bibfield  {journal}
  {\bibinfo  {journal} {Am. J. Phys.}\ }\textbf {\bibinfo {volume} {66}},\
  \bibinfo {pages} {867} (\bibinfo {year} {1998})}\BibitemShut {NoStop}%
\bibitem [{\citenamefont {Lekner}(2007)}]{Lekner_AJP07}%
  \BibitemOpen
  \bibfield  {author} {\bibinfo {author} {\bibfnamefont {J.}~\bibnamefont
  {Lekner}},\ }\href {https://doi.org/10.1119/1.2787015} {\bibfield  {journal}
  {\bibinfo  {journal} {Am. J. Phys.}\ }\textbf {\bibinfo {volume} {75}},\
  \bibinfo {pages} {1151} (\bibinfo {year} {2007})}\BibitemShut {NoStop}%
\bibitem [{\citenamefont {\c{S}akiro\u{g}lu}\ \emph {et~al.}(2012)\citenamefont
  {\c{S}akiro\u{g}lu}, \citenamefont {Ungan}, \citenamefont {Yesilgul},
  \citenamefont {Mora-Ramos}, \citenamefont {Duque}, \citenamefont {Kasapoglu},
  \citenamefont {Sari},\ and\ \citenamefont {S{\"{o}}kmen}}]{Sakiroglu_PLA12}%
  \BibitemOpen
  \bibfield  {author} {\bibinfo {author} {\bibfnamefont {S.}~\bibnamefont
  {\c{S}akiro\u{g}lu}}, \bibinfo {author} {\bibfnamefont {F.}~\bibnamefont
  {Ungan}}, \bibinfo {author} {\bibfnamefont {U.}~\bibnamefont {Yesilgul}},
  \bibinfo {author} {\bibfnamefont {M.}~\bibnamefont {Mora-Ramos}}, \bibinfo
  {author} {\bibfnamefont {C.}~\bibnamefont {Duque}}, \bibinfo {author}
  {\bibfnamefont {E.}~\bibnamefont {Kasapoglu}}, \bibinfo {author}
  {\bibfnamefont {H.}~\bibnamefont {Sari}},\ and\ \bibinfo {author}
  {\bibfnamefont {Ä.}~\bibnamefont {S{\"{o}}kmen}},\ }\href
  {https://doi.org/10.1016/j.physleta.2012.04.028} {\bibfield  {journal}
  {\bibinfo  {journal} {Phys. Lett. A}\ }\textbf {\bibinfo {volume} {376}},\
  \bibinfo {pages} {1875} (\bibinfo {year} {2012})}\BibitemShut {NoStop}%
\bibitem [{\citenamefont {Radovanovi{\'{c}}}\ \emph {et~al.}(2000)\citenamefont
  {Radovanovi{\'{c}}}, \citenamefont {Milanovi{\'{c}}}, \citenamefont
  {Ikoni{\'{c}}},\ and\ \citenamefont {Indjin}}]{Radovanovic_PLA00}%
  \BibitemOpen
  \bibfield  {author} {\bibinfo {author} {\bibfnamefont {J.}~\bibnamefont
  {Radovanovi{\'{c}}}}, \bibinfo {author} {\bibfnamefont {V.}~\bibnamefont
  {Milanovi{\'{c}}}}, \bibinfo {author} {\bibfnamefont {Z.}~\bibnamefont
  {Ikoni{\'{c}}}},\ and\ \bibinfo {author} {\bibfnamefont {D.}~\bibnamefont
  {Indjin}},\ }\href {https://doi.org/10.1016/S0375-9601(00)00238-3} {\bibfield
   {journal} {\bibinfo  {journal} {Phys. Lett. A}\ }\textbf {\bibinfo {volume}
  {269}},\ \bibinfo {pages} {179} (\bibinfo {year} {2000})}\BibitemShut
  {NoStop}%
\bibitem [{\citenamefont {Y\i{}ld\i{}r\i{}m}\ and\ \citenamefont
  {Tomak}(2005)}]{Yildirim_PRB05}%
  \BibitemOpen
  \bibfield  {author} {\bibinfo {author} {\bibfnamefont {H.}~\bibnamefont
  {Y\i{}ld\i{}r\i{}m}}\ and\ \bibinfo {author} {\bibfnamefont {M.}~\bibnamefont
  {Tomak}},\ }\href {https://doi.org/10.1103/PhysRevB.72.115340} {\bibfield
  {journal} {\bibinfo  {journal} {Phys. Rev. B}\ }\textbf {\bibinfo {volume}
  {72}},\ \bibinfo {pages} {115340} (\bibinfo {year} {2005})}\BibitemShut
  {NoStop}%
\bibitem [{\citenamefont {Mazzarella}\ \emph {et~al.}(2010)\citenamefont
  {Mazzarella}, \citenamefont {Moratti}, \citenamefont {Salasnich},\ and\
  \citenamefont {Toigo}}]{MAZZARELA_JPB10}%
  \BibitemOpen
  \bibfield  {author} {\bibinfo {author} {\bibfnamefont {G.}~\bibnamefont
  {Mazzarella}}, \bibinfo {author} {\bibfnamefont {M.}~\bibnamefont {Moratti}},
  \bibinfo {author} {\bibfnamefont {L.}~\bibnamefont {Salasnich}},\ and\
  \bibinfo {author} {\bibfnamefont {F.}~\bibnamefont {Toigo}},\ }\href
  {https://doi.org/10.1088/0953-4075/43/6/065303} {\bibfield  {journal}
  {\bibinfo  {journal} {J. Phys. B At. Mol. Opt. Phys.}\ }\textbf {\bibinfo
  {volume} {43}},\ \bibinfo {pages} {065303} (\bibinfo {year}
  {2010})}\BibitemShut {NoStop}%
\bibitem [{\citenamefont {Mazzarella}\ and\ \citenamefont
  {Salasnich}(2010)}]{Mazzarella_PRA10}%
  \BibitemOpen
  \bibfield  {author} {\bibinfo {author} {\bibfnamefont {G.}~\bibnamefont
  {Mazzarella}}\ and\ \bibinfo {author} {\bibfnamefont {L.}~\bibnamefont
  {Salasnich}},\ }\href {https://doi.org/10.1103/PhysRevA.82.033611} {\bibfield
   {journal} {\bibinfo  {journal} {Phys. Rev. A}\ }\textbf {\bibinfo {volume}
  {82}},\ \bibinfo {pages} {033611} (\bibinfo {year} {2010})}\BibitemShut
  {NoStop}%
\bibitem [{\citenamefont {Yang}(2010)}]{Yang_10}%
  \BibitemOpen
  \bibfield  {author} {\bibinfo {author} {\bibfnamefont {J.}~\bibnamefont
  {Yang}},\ }\href {https://doi.org/10.1137/1.9780898719680} {\emph {\bibinfo
  {title} {{Nonlinear Waves in Integrable and Nonintegrable Systems}}}}\
  (\bibinfo  {publisher} {Society for Industrial and Applied Mathematics},\
  \bibinfo {year} {2010})\BibitemShut {NoStop}%
\end{thebibliography}%

\end{document}